\documentclass[12pt]{article}
\usepackage{epsf}

\newcommand{\beq}{\begin{equation}}
\newcommand{\eeq}{\end{equation}}
\newcommand{\ba}{\begin{array}}
\newcommand{\ea}{\end{array}}
\newcommand{\beqa}{\begin{eqnarray}}
\newcommand{\eeqa}{\end{eqnarray}}
\newcommand{\lsim}{\stackrel{<}{_\sim}}

\begin{document}

\thispagestyle{empty}
\setcounter{page}{0}
\begin{flushright}
SLAC-PUB-8752\\
hep-ph/0108074 \\
December 2001
\end{flushright}
\vspace*{1.5cm}
\centerline{\Large\bf Probing for New Physics in Polarized $\Lambda_b$
decays at the $Z$
}
\vspace*{2cm}
\centerline{ Gudrun Hiller${}^a$\footnote{Work supported by
Department of Energy contract DE--AC03--76SF00515.} and Alex 
Kagan${}^{b}$\footnote{Work supported by
the Department of Energy Grant No. DE-FG02-84ER40153.}}
\bigskip
\centerline{\sl ${}^a$Stanford Linear Accelerator Center,
                Stanford University, Stanford, CA 94309, USA }
\centerline{\sl ${}^b$ Department of Physics, University of Cincinnati,
Cincinnati, OH 45221, USA}

\vspace*{1.5cm}
\centerline{\bf Abstract}
\vspace*{0.3cm}
\noindent

Polarized $\Lambda_b \to \Lambda \gamma$ decays
at the $Z$ pole are shown to be
well suited for probing a large variety of New Physics effects.
A new observable is proposed, the angular asymmetry between the
$\Lambda_b$ spin and photon momentum, which is
sensitive to the relative strengths of the opposite chirality and
Standard Model
chirality $b \to s \gamma$ dipole operators.
Combination with the $\Lambda $ decay polarization asymmetry
and comparison with the $\Lambda_b$ polarization extracted from
semileptonic decays
allows important tests of the $V-A$ structure of the Standard Model.
Modifications of the rates and angular asymmetries
which arise at next-to-leading order are discussed.
Measurements for
$\Lambda_b \to \Lambda \gamma$ and the CP conjugate
mode, with branching ratios of a few times $10^{-5}$, are shown to
be sensitive to non-standard
sources of CP violation in the $\Lambda_b \to \Lambda \gamma$
matrix element.
Form factor relations for heavy-to-light baryon decays
are derived in the large energy limit, which are of general interest.

\vfill

\newpage
\pagenumbering{arabic}

\section{Introduction}

Flavor changing neutral current (FCNC) $b$-decays provide
important tests of the Standard Model (SM) at the quantum level and,
at the same time, place severe constraints on New Physics extensions.
In this paper we investigate the possibility of searching for
New Physics in radiative FCNC decays induced by $b \to s \gamma$ transitions.
The relevant low energy effective Hamiltonian at
leading-order (LO) in $\alpha_s$
is given by \cite{heffbig}
\begin{equation}
{\cal{H}}_{eff} = - {G_F \over \sqrt{2} } V_{t s}^\ast  V_{tb}\,
\left[ C_{7}
Q_{7 } + C^\prime_{7} Q^\prime_{7}\,\right],
\end{equation}
with the electromagnetic dipole operators $Q_{7},~Q^\prime_{7}$ written as
\beq
Q_{7}           = \frac{e}{8 \pi^2} m_b \bar{s} \sigma_{\mu \nu}
                     R b F^{\mu \nu}~,\qquad
Q^\prime_{7}    = \frac{e}{8 \pi^2} m_b \bar{s} \sigma_{\mu \nu}
                      L b F^{\mu \nu}.
\eeq
Here $L \equiv 1 - \gamma_5$ and $R \equiv 1 + \gamma_5 $ are
proportional to the left-
and right-handed projectors.  Renormalization scale dependence of the 
Wilson coefficients $C_i$ and operator
matrix elements is understood.
In the SM, the contribution to $Q^\prime_{7}$ is suppressed with respect to
the one to
$Q_{7}$ by the small mass-insertion along the external
s-quark line and is usually neglected, i.e.,
$C_{7  \, \rm{SM}}^\prime = m_s/m_b C_{7 \, \rm{SM}}$.
However, in many extensions of the SM new contributions to $C^\prime_{7}$
are not necessarily suppressed and can be comparable to
$C_{7 \, \rm{SM}}$ since the requisite helicity-flip is
along a massive fermion propagator inside the loop.
Examples are Left-Right symmetric models,
supersymmetric models with large
left-right squark flavor mixing, and models containing new vectorlike
quarks.

The branching fraction for inclusive $B \to X_s \gamma $ decays
has been measured \cite{cleobsg,lepbsg,bellebsg} and is
consistent with the SM prediction,
e.g.~\cite{misiak,GHW,kaganneubert,misiakgambino}.
The measurement constrains the combination
${\cal{B}}(B \to X_s \gamma)
\propto |C_{7}|^2+|C_{7}^\prime|^2
\simeq |C_{7\, \rm{SM}}|^2$,
which is a circle in the $C_{7}$-$C_{7}^\prime$ plane.
Thus, complementary data are needed for a model-independent
determination of
$C_{7}$ and $C_{7}^\prime$ separately.
One suggestion has been to probe the photon helicity
via the mixing induced CP-asymmetry in neutral $B_{(d,s)} \to M \gamma$
decays, where $M=\omega,\rho,K^*,\phi$ \cite{atwoodgronau}.
Other methods have aimed at analyzing the angular distribution of the
subsequent decay products.
These include correlation studies in the dilepton mode
$B \to K^* (\to K \pi) \gamma^* (\to \ell^+ \ell^-)$ in the
low dilepton mass region \cite{frombsee,yuvipirjol}, and radiative
$B$-decays into excited kaons yielding
$K \pi \pi^0 \gamma$ final states \cite{yuvinew}.

We propose here to probe the ratio $C_{7}^\prime / C_{7}$ in
polarized $\Lambda_b \to \Lambda \gamma$ decays by measuring the angular
asymmetry between the $\Lambda_b$ spin and the momentum of the photon
(or $\Lambda$).
The longitudinal polarization of $\Lambda_b$ baryons produced in $Z$ decays
has been measured in semileptonic $ \Lambda_b \to \Lambda_c \ell \nu_{\ell} X$
decays and is found to retain a sizable fraction of the parent
$b$-quark polarization
\cite{bonvicinirandall,aleph96,opal98,delphi99}.
Besides the angular asymmetry, which
explicitly makes use of the polarization feature of the $\Lambda_b$ baryons,
a second `helicity{'} observable can be used to probe
the quark chiralities:
the $\Lambda$ polarization variable
associated with the secondary decays $\Lambda \to p \pi^- $, first
proposed in \cite{mannelrecksiegel}
for unpolarized $\Lambda_b \to \Lambda \gamma$ decays.
Because these two observables are independent, as we will show,
their measurements
allow consistency checks and their combined analysis
greatly increases the New Physics reach.
We rederive the $\Lambda$ decay polarization asymmetry and find an
expression which differs from previous ones obtained in the literature
\cite{mannelrecksiegel,huangyan98,chuahehou}.

{}From a general Lorenz decomposition it follows that only a
single overall hadronic form factor $F (0)$ enters the
$\Lambda_b \to \Lambda \gamma$ amplitude,
and therefore it cancels in the forward-backward asymmetries.
Based on studies of $\Lambda_b \to \Lambda \gamma $ \cite{mannelrecksiegel}
and $B \to K^* \gamma $ decays
\cite{LDkstar}-\cite{desh}
corrections of at most a few percent can be expected from
long-distance interactions
so there is very little hadronic
uncertainty in the SM prediction for the helicity observables.

Heavy quark effective theory (HQET) spin symmetry
arguments applied to heavy-to-light baryon form factors
\cite{mannelrecksiegel,mannelrobertshussain91} relate
the overall form factor $F (0) $ entering the
$\Lambda_b \to \Lambda \gamma $ matrix element
to two universal form factors $F_1 (0) $ and $F_2 (0)$.
Consistent estimates for $F_1 (0) $ have been obtained
from data on semileptonic $\Lambda_c $ decays  \cite{mannelrecksiegel}
and from QCD sum rules \cite{huangyan98}.
We find that a new application of large energy effective
theory (LEET) \cite{charlesetal} to heavy-to-light baryon form factors
fixes the ratio $F_2 (0) / F_1 (0) $,
which allows us to use the information on $F_1 (0) $ to estimate
$F(0)$ and therefore the total $\Lambda_b \to \Lambda \gamma $ rate.

At next-to-leading order (NLO) in $\alpha_s$
direct CP violation can be probed in $b \to s \gamma $ mediated decays.
We estimate the dominant NLO effects in the
$\Lambda_b \to \Lambda \gamma $ matrix element
and allow for non-standard CP violation in contributions to both the
SM and opposite chirality dipole operators.
Rates and helicity observables for the untagged (CP averaged) and
flavor tagged cases are worked out.
Experimental discrimination between the CP
conjugate decays is easy because they are self tagging.

It should be stressed that
while other proposals for probing the ratio
$C_{7}^\prime / C_{7} $
\cite{atwoodgronau,frombsee,yuvipirjol,yuvinew,mannelrecksiegel}
can be carried out at upgraded $e^+ e^-$ $B$-factories
or at hadron colliders, the
angular asymmetry observable using initial state polarization is unique to a
high luminosity $e^+ e^-$ machine running at the $Z$ pole.
Proposals exist for a so-called GigaZ option with
$2 \cdot 10^9$ $Z$ bosons per year \cite{tesla,usaLC},
corresponding to approximately $3.5 \cdot 10^7$ $b$-flavored baryon
decays.
For recent discussions of the $b$ physics potential at a $Z$ factory
see \cite{batgigaZ}.
With a branching fraction estimate
${\cal{B}}(\Lambda_b \to \Lambda \gamma) \simeq 7.5 \cdot 10^{-5}$
we expect approximately $2600 $ exclusive
$\Lambda_b \to \Lambda \gamma$ decays per year.

This paper is organized as follows: In Section \ref{sec:ff} we discuss form
factor relations for $\Lambda_b \rightarrow \Lambda $
transitions following from HQET/LEET.
In Section \ref{sec:asymmetry} we define
the two angular asymmetry observables for $\Lambda_b \to \Lambda \gamma$
and study
their sensitivities, separately and combined, to the ratio
$C_{7}^\prime / C_{7} $.
Section \ref{sec:NLO} is devoted to a discussion of next-to-leading
order effects including CP violation.
In Section \ref{sec:conclusions} we conclude and give a brief outlook on
further opportunities in $b$-physics at hadron colliders and at the GigaZ.

\section{Form Factor Preliminaries \label{sec:ff}}

The most general decomposition of $\Lambda_b \to \Lambda$
matrix elements for the dipole transition into an on-shell photon is given by
\beq
\label{eq:sigma}
\langle\Lambda(p^\prime,s^\prime) | \bar{s} \sigma_{\mu \nu} (1 \pm
\gamma_5) q^\nu b
|\Lambda_b(p,s)\rangle  =
F(0) \bar{u}_\Lambda \sigma_{\mu \nu}(1 \pm \gamma_5) q^\nu u_{\Lambda_b},
\eeq
where $p^{(')}$ and $s^{(')}$ denote the baryon momenta and spins,
respectively,
$q = p-p^\prime $, and
$u_{\Lambda}, u_{\Lambda_b}$ are the baryon spinors.
We stress that only one overall form factor $F(0)$ enters
(this follows from the identity
$\sigma^{\mu \nu} \gamma_5=\frac{i}{2} \epsilon^{\mu \nu \rho \sigma}
\sigma_{\rho \sigma}$), so that the different helicities do not mix.

In the following we work out form factor relations
for heavy-to-light baryon decays
which follow from certain limits.
This provides a realization of the physical picture of helicity conservation,
and allows us to estimate $F(0)$ and therefore the
total $\Lambda_b \to \Lambda \gamma $ rate
in terms of existing form factor calculations and measurements.

\subsection{The Large Energy and Heavy Quark Limits}

In the decays under consideration
a baryon containing a
heavy quark decays into a light baryon with small $q^2 \ll m_{\Lambda_b}^2$.
In these heavy-to-light decays the energy $E$
of the light baryon
$E=(m_{\Lambda_b}^2+m_\Lambda^2-q^2)/2 m_{\Lambda_b} $ in the
parent baryon{'}s rest frame is large compared to
the strong interaction scale and the light quark or baryon masses.
This is precisely the kinematical situation for which one can consider
the Large Energy Effective Theory \cite{charlesetal},
originally introduced in Ref.~\cite{dugangrinstein}.
It arises from a
systematic $1/E$ expansion of the QCD Lagrangian of the final active
light quark.
Neglecting hard interactions with the spectators and other soft degrees
of freedom, the momenta of the final active
quark, $p^\prime_{\rm{quark}}$, and the final hadron, $p^\prime $, are equal
modulo a small
residual momentum $k \simeq \Lambda_{QCD}$:
$p^\prime_{\rm{quark} \; \mu}=E n_\mu+k_\mu$, where
$n \equiv p^{\prime}/E$.  At leading order in LEET $n$ is light-like
($n^2 = 0$), i.e., terms of order $m_\Lambda^2 /E^2 $ are
neglected, and the final LEET quark is on-shell with $\not{n} s=0$.
For details we refer the reader to \cite{charlesetal}.

The assumption of soft contribution dominance in LEET is consistent
with an HQET
description of the initial decaying $b$ quark.
Symmetries which arise in the combined LEET/HQET limit
imply relations among form factors for heavy-to-light decays.
They will receive corrections at order $1/m_b$, $1/E$ and $\alpha_s$.
For $B$-meson decays into a light pseudoscalar or vector meson, the
leading order form factor relations have
been worked out in \cite{charlesetal}.
Perturbative ${\cal O}(\alpha_s )$ vertex and hard scattering corrections have
been found to typically lie below the
$10 \%$ level \cite{benekefeldmann00}.
The soft parts of the form factor
relations, found in \cite{charlesetal}, have been confirmed in
'collinear-soft' effective theory \cite{colleff}.

The $\alpha_s (\sqrt{m_b \Lambda_{QCD} }) $ suppression of hard
scattering form factor contributions in heavy-to-light $B$-meson decays
\cite{benekefeldmann00} supports the starting assumption of
soft dominance and the applicability of HQET in this regime.
We will assume that this suppression
also holds for heavy-to-light $b$-baryon
decays so that a perturbative expansion in $1/m_b$, $1/E$ and $\alpha_s$
is again sensible.
A rigorous treatment of higher-order corrections to heavy-to-light
baryon form factors is beyond the scope
of this paper and is left for future work.
We will however briefly comment on $1/m_b$ corrections below.

\subsection{LEET/HQET Form Factor Relations}

Heavy quark spin symmetry
implies the following parametrization of hadronic
matrix elements \cite{mannelrobertshussain91}
in the $m_b \to \infty$ limit
\beqa
\langle \Lambda(p^\prime,s^\prime) | \bar{s} \Gamma b |\Lambda_b(p,s) \rangle =
\bar{u}_\Lambda \left( F_1(q^2)+\not{v} F_2 (q^2) \right) \Gamma u_{\Lambda_b},
\label{eq:hqetff}
\eeqa
which involves only two universal form factors
for any Dirac structure $\Gamma$.
This yields for example for $\Gamma=\gamma_\mu$
\beqa
<\Lambda(p^\prime,s^\prime) | \bar{s} \gamma_\mu b |\Lambda_b(p,s)>=
\bar{u}_\Lambda \left[ ( F_1(q^2)- F_2 (q^2)) \gamma_\mu+2  F_2 (q^2) v_\mu
\right] u_{\Lambda_b},
\label{eq:hqetffvector}
\eeqa
where $v=p/m_{\Lambda_b}$ denotes the velocity of the heavy
baryon.

Comparing Eq.~(\ref{eq:sigma}) with Eq.~(\ref{eq:hqetff}) for the dipole
transition and
using the HQET relation $\not{v} b = b$ yields
\beqa
F (0) = F_1 (0) + \frac{m_\Lambda}{m_{\Lambda_b}} F_2 (0).
\label{eq:hqetdipole}
\eeqa

It is apparent from Eq.~(\ref{eq:hqetff}) that the helicity of the
$\Lambda_b$ is determined by the helicity of the heavy $b$-quark, and
that the light degrees of freedom in the $\Lambda_b$ are in a spin-0 state.
This is what one would expect in the naive valence quark picture
of hadrons, or diquark picture of baryons.
However, in general the correspondence between the helicity of the
active light quark and
the helicity of the light baryon is broken by the ratio $F_2/F_1$.

LEET allows us to relate the two form factors $F_1$ and $F_2$.
Contracting the 4-vector $n_\mu$ with the matrix element over the vector
current given in Eq.~(\ref{eq:hqetffvector}) and using
$\not{n} s = 0$, $v.n=1$  we derive
at lowest order in LEET/HQET
\begin{equation}
\label{eq:leet}
F_2(E,m_b)/F_1(E,m_b)=-\frac{m_\Lambda}{2 E},
\end{equation}
where the dependence on the expansion parameters has been made explicit.
This is in concordance with the physical picture
for heavy-to-light $B$-meson decays recently obtained in
Ref.~\cite{gg}:
The helicity of the active light quark is 'inherited' by the final hadron.
Corrections to this are proportional to light masses and are
suppressed by $1/E$.  We estimate their size to be less than
$m_\Lambda /m_{\Lambda_b} \sim 20 \%$.
Note that Eq.~(\ref{eq:leet}) holds at lowest order in
collinear-soft effective theory \cite{colleff}.

Heavy quark relations like Eq.~(\ref{eq:hqetff}) receive
$1/m_b$ corrections, which are small near zero
recoil $q^2 \approx q^2_{max}$ since there is little
energy transfer to the light degrees of freedom.
Near maximal recoil one might think that the light degrees of freedom
could receive large excitations so that this is no longer the case.
However, in the $E\to \infty$ limit the LEET/HQET effective theory
is independent of the light hadron energy, $E$,
not just the heavy quark mass $m_b$.
The LEET light quark field, in particular, only depends on a 
`residual{'} momentum
of order $\bar{\Lambda} $. The soft form factor contribution 
dominance assumption,
which requires that production of the light hadron at low $q^2$ is
governed by the end-point region of its
wave function, is used to justify the applicability of LEET/HQET
to heavy-to-light decays in this kinematical regime.
It implies that $1/m_b $, 1/E
and perturbative $\alpha_s$ corrections
to form factor relations such as Eq.~(\ref{eq:hqetff}) remain small 
and well defined.

We briefly comment on implications of LEET for
$1/m_b$ corrections to heavy-to-light baryon form factors 
at large recoil.
To facilitate the discussion 
we introduce the general decomposition for the vector current 
\beqa
\langle \Lambda(p^\prime,s^\prime) | \bar{s} \gamma_\mu b 
|\Lambda_b(p,s) \rangle = \bar{u}_\Lambda 
\left[ V_1(q^2) \gamma_\mu +V_2(q^2) v_\mu+ V_3(q^2) n_\mu \right] 
 u_{\Lambda_b},
\label{eq:hqetv}
\eeqa
where at leading order, by comparison with Eq.~(\ref{eq:hqetffvector}),
\beqa
V_1(q^2)=F_1(q^2)-F_2(q^2), ~~~~~~~~~~~~ V_2(q^2)= 2 F_2(q^2), 
~~~~~~~~~~~~ V_3(q^2)=0.
\eeqa
In HQET,  additional  non
perturbative form factors are introduced at order $1/m_b$
\cite{datta}, which lead to shifts in the $V_i$.
The use of LEET leads to relations among the new form factors entering
at order $1/m_b$.  
Remarkably, they imply that neglecting radiative corrections 
the leading order relation
\beqa
V_2(E,m_b)/V_1(E,m_b)=2 F_2(E,m_b)/F_1(E,m_b)=-\frac{m_\Lambda}{E},
\eeqa
remains unchanged. An analogous result holds for the corresponding 
axial vector current form factors.
We also find, that the infinite $m_b$ relation 
$F(0)=F_1(0) + {\cal{O}}(m_\Lambda^2/E^2)$, see Eq.~(\ref{eq:hqetdipole})
is modified so that
$F (0) = F_1 (0) (1 + {\cal{O}}(\bar{\Lambda}/m_b) )$.

The form factor relations apply generally to
any $b \to q$ mediated heavy-to-light baryon decay, where $q=u,d,s$.
Examples are $\Lambda_b \to p \ell^- \bar{\nu}_\ell$
which is sensitive to $V_{ub}$, rare $\Lambda_b \to \Lambda +X$
decays like the one
under consideration, and their CKM
suppressed counterparts $\Lambda_b \to n +X$, where
$X=\gamma, \ell^+ \ell^-, \nu \bar{\nu}$, etc.
Note that the flavor dependence of the ratio $F_2 /F_1$ in
Eq.~(\ref{eq:leet}) is small, since the light
baryon mass differences are small compared to $m_{\Lambda_b}$.
However, it indicates that $F_2 /F_1$ decreases for lighter final
state baryons.

It is interesting to compare the LEET prediction in Eq. (\ref{eq:leet})
with other determinations of the form factor ratio.
For the
radiative decay $\Lambda_b \to \Lambda \gamma$,
with $E=2.9$ GeV, we obtain the LEET ratio
\begin{equation}
\label{eq:leetgamma}
F_2 (0)/ F_1 (0) = -0.19.
\end{equation}
This agrees well with a QCD sum rule calculation \cite{huangyan98}, which gives
$F_2(0)/F_1(0)=-0.20 \pm 0.06$.
For $\Lambda_b \to p$ we obtain $F_2(0)/F_1(0)|_p =-0.16$, which is
also consistent
with the QCD sum rule result
$F_2(0)/F_1(0)|_p =-0.18 \pm 0.07$ \cite{lambdatop}.

For charmed baryons there exists a CLEO measurement
of this ratio coming from semileptonic
$\Lambda^+_c \to \Lambda e^+ \nu_e$
decays,
$\langle F_2/F_1 \rangle^{\rm{data}}_c = -0.25 \pm 0.14 \pm 0.08$, where
the flavor of the decaying heavy quark, and an average over phase
space are
indicated \cite{cleolambdac}.
Although naively we do not expect LEET to be applicable to charm
decays since the maximal hadronic energy is not much
larger than $m_\Lambda $, it is
interesting to note  that the LEET/HQET prediction
$<F_2/F_1>_c^{\rm{theory}}=-0.44$
agrees with the CLEO result
in sign and size at the $1\sigma$ level.  Note that we have evaluated
Eq.~(\ref{eq:leet}) at the average value of
$q^2$, $\langle q^2 \rangle =0.7 \mbox{GeV}^2$.

The authors of Ref.~\cite{mannelrecksiegel} have used the same CLEO
data on semileptonic $\Lambda_c$ decays together with Eq.~(\ref{eq:hqetff})
to obtain $F_1(0)=0.22$ (dipole) and $F_1(0)=0.45$
(monopole) for $\Lambda_b $ decays.  As indicated, this requires an
assumption about the $q^2$ dependence of
the form factors in order to extrapolate from charm to bottom decays
which leads to large theoretical uncertainties \cite{gg}.
However, the latter (monopole) value of $F_1(0)$ is in reasonable
agreement with
$F_1(0)=0.50 \pm 0.03$, derived from QCD sum rules \cite{huangyan98}.
Noting that to leading order in HQET/LEET $F(0)=F_1(0)$,
we choose $F(0)=0.50$ 
to estimate the normalization of the decays under investigation.
We recall that the dependence on the form factor
drops out in the angular asymmetry observables.

We briefly mention an interesting application of our
results for heavy-to-light baryon form factors at large recoil.
In Ref.~\cite{Chen:2001sj}, to which we refer the reader for details,
it has been empirically observed that the position
of the zero of the dilepton forward-backward asymmetry
in $\Lambda_b \rightarrow
\Lambda \ell^+ \ell^-$ decays parametrically has
very little dependence on the form factors.
We argue that this is a consequence of LEET:
corrections to the universal zero in inclusive $b \to s \ell^+ \ell^-$
decays are proportional to $m_\Lambda^2/E^2$ and
$m_\Lambda/E ~F_2/F_1$, which are of higher order in LEET.

\section{Angular Asymmetry in $\Lambda_b \to \Lambda \gamma$ and New
Physics \label{sec:asymmetry}}

The ratio $r \equiv C_{7}^\prime / C_{7} $ can be probed by
looking at the angular distributions of
the spin degrees of freedom with respect to the photon (or $\Lambda $)
momentum vector in $\Lambda_b \to \Lambda \gamma$ decays.
At the $Z$ both initial and final baryons will be polarized.
We therefore begin by giving the
differential decay width
with the dependence on both baryon spins included.
Using Eq.~(\ref{eq:sigma}) we obtain the exact LO result,
which is in agreement with the corresponding expression for
$Baryon \to Baryon+Vector$ decays derived in \cite{Pakvasa:1990if} in the limit
of a massless transverse vector state
\beq
\label{eq:photonsS}
d \Gamma (\Lambda_b \to \Lambda \gamma) =
\Gamma_0 |C_{7}|^2 \frac{d \Omega_S}{4 \pi} \frac{d \Omega_s}{4\pi}
\left( (1 + |r|^2) [1
- (\vec{S} \cdot \hat{p}_\Lambda ) (\vec{s} \cdot \hat{p}_\Lambda )] +
(1-|r|^2 ) [\vec{S} \cdot \hat{p}_\Lambda - \vec{s} \cdot
\hat{p}_\Lambda ] \right).
\eeq
Here,
$\vec{S}$ and $\vec{s}$ are unit vectors parallel to the spins of the
$\Lambda_b $ and $\Lambda $ in their respective rest frames,
$\Omega_S $ and $\Omega_s $ are their solid angle elements,
$\hat{p}_\Lambda $ is a unit
vector pointing in the direction of the $\Lambda $ momentum, and
\beq
\label{eq:Gamma0}
\Gamma_0 \equiv \frac{\alpha G_F^2 |V_{tb} V_{ts}^*|^2}{32 \pi^4}
m_{\Lambda_b}^3
   m_b^2 \left(1- \frac{m_\Lambda^2}{m_{\Lambda_b}^2} \right)^3
|F(0)|^2  .
\eeq
The total decay rate is $\Gamma  =\Gamma_0 |C_{7}|^2 (1+|r|^2)$,
and our
estimate for the branching fraction
is \cite{pdg00}
\beq
{\cal{B}} (\Lambda_b \to \Lambda \gamma) =
\frac{1.23 \mbox{ps}}{\tau(\Lambda_b)}
\left(\frac{m_b}{4.4 \mbox{GeV}} \right)^2
\left|\frac{V_{tb} V_{ts}^*}{0.04} \right|^2
\left|\frac{F(0)}{0.5} \right|^2
\left| \frac{C_{7}}{-0.31} \right|^2
(1+|r|^2)  \times 7.9 \cdot 10^{-5}.
\eeq
Taking $F (0) \approx 0.5$, as discussed in the previous section and
$|C_{7} |^2 + |C_{7}^\prime |^2 \approx |C_{7 \, \rm{SM}}|^2 $,
the SM branching fraction can be expected to lie in the range
$(3-10)  \cdot 10^{-5}$.

Note that there are long distance effects due to intermediate 
$c \bar{c}$ states which can lead to
small helicity changing contributions.
A model independent $\Lambda_{QCD} /m_c$ expansion 
has been performed for both inclusive $b \to s \gamma$ 
\cite{buchallaligetigrantvoloshin} and exclusive
$B \to K^* \gamma$ decays \cite{LDkstar}, 
yielding contributions
which are only a few percent of the short-distance amplitudes.
Resonance exchange models 
making use of photoproduction data to evaluate the
charmonium couplings at the right kinematical point \cite{desh}
are consistent with the $1/m_c$ expansion.
A model calculation for $\Lambda_b \to \Lambda \gamma$ decays
\cite{mannelrecksiegel}
based on \cite{desh} again yields 
contributions at the few percent level.
Cabbibo suppressed internal $W$
exchange has also been found to contribute
at the percent level to both $B \to K^* \gamma $ \cite{pirjol}
and $\Lambda_b \to \Lambda \gamma$ decays \cite{mannelrecksiegel}.
As the overall long distance uncertainties turn out to be well below 
the experimental sensitivity, see Table \ref{tab:ranges}, 
they will be neglected in this work.

We now introduce our observable, the angular asymmetry for
polarized $\Lambda_b$ baryons.
We define $\theta_S $ as the angle between
$\vec{S}$ and $\hat{p}_\Lambda $.
Starting from Eq.~(\ref{eq:photonsS})
it is straightforward to obtain the forward-backward asymmetry
${\cal{A}}_{\theta_S}$,
\beq
\label{eq:AthetaS}
{\cal{A}}_{\theta_S} \equiv
\frac{1}{\Gamma}
\left( {\displaystyle \int_{0}^{1} d \cos \theta_S \frac{d \Gamma}{d \cos
\theta_S }-
\int_{-1}^{0} d \cos \theta_S \frac{d \Gamma}{ d \cos \theta_S } }\right)
= {1\over 2} \frac{1-|r|^2}{1+|r|^2}.
\eeq
The polarization $P_{\Lambda_b}$ of $\Lambda_b$ baryons produced in
$Z$ decays then
gives us the angular asymmetry observable, ${\cal{A}}^{\gamma}$,
defined (in the $\Lambda_b$ rest frame) as the forward-backward asymmetry of
the photon momentum with respect to the $\Lambda_b$ boost axis,
\beq
{\cal{A}}^\gamma \equiv -P_{\Lambda_b} {\cal{A}}_{\theta_S}
= - {P_{\Lambda_b} \over 2}
\frac{1-|r|^2}{1+|r|^2}. \label{eq:alambdab}
\eeq
For $r \ll 1$, as in the SM, small angles
$\theta_S \simeq 0$ are favored and the photon is
emitted back-to-back with respect to the spin of the $\Lambda_b$, or
preferentially parallel to the boost axis since $P_{\Lambda_b} < 0$.

To make contact with experiment we relate
${\cal{A}}^\gamma $ to the average
longitudinal momentum of the photon with respect to the $\Lambda_b $
boost axis,
$<q_{\parallel}^*> =2/3 E_\gamma^* {\cal{A}}_{\gamma}$,
where
$E_{\gamma}^*=(m_{\Lambda_b}^2-m_{\Lambda}^2)/(2 m_{\Lambda_b})=2.7$ GeV
is the photon energy
(starred quantities are in the $\Lambda_b$ rest frame, unstarred
quantities are in the lab frame).
Finally, we arrive at an expression for the average longitudinal
momentum $<q_{\parallel}>_{\beta}$ of the photon in the lab frame
with respect to the boost axis for a fixed
boost $\beta = |\vec{p}_{\Lambda_b} | /E_{\Lambda_b}$,
\beq
<q_{\parallel}>_{\beta}=\gamma ( \beta E_\gamma^* + <q_{\parallel}^*> )=
\gamma  E_\gamma^* (\beta + {2\over{3}} {\cal{A}}_{\gamma}),
\eeq
which allows the extraction of ${\cal{A}}_{\gamma}$.

The sensitivity of ${\cal{A}}^\gamma $ to New Physics effects depends on the
magnitude of the $\Lambda_b$ polarization.
In the heavy quark limit, $\Lambda_b${'}s produced in $Z$ decays pick
up the (longitudinal) polarization of the $b$-quark, $P_b=-0.94 $
for $\sin^2 \theta_W=0.23$.
Depolarization effects during the fragmentation process were studied in
Ref.~\cite{peskinfalk}.  Based on HQET and poorly
known non-perturbative parameters extracted from data, the
average longitudinal $\Lambda_b$ polarization was estimated to be
$P_{\Lambda_b}^{\rm{HQET}}=-(0.69 \pm 0.06)$.
We will instead use the central value of the OPAL
Collaboration's measurement,
$P_{\Lambda_b}=-0.56^{+0.20}_{-0.13} \pm 0.09$ \cite{opal98}, as an
input in our analysis.
The LEP measurements of
$P_{\Lambda_b}$ \cite{aleph96,opal98,delphi99} are obtained from the
lepton spectra in semileptonic
$\Lambda_b \to \Lambda_c \ell \nu_{\ell} X$ decays, assuming purely
SM $V-A$ currents \cite{bonvicinirandall}.
With a few times $10^2$ more events at a GigaZ machine the error
should decrease substantially.
This issue certainly deserves further study.

Next we discuss the second 'helicity' observable which follows from
a spin analysis of the final baryon.
The $\Lambda $ polarization
variable $\alpha_\Lambda $ is defined in the
differential decay width as \cite{pdg00}
$d \Gamma /d\Omega_s \propto (1 + \alpha_\Lambda \vec{s} \cdot
\hat{p}_\Lambda )$.  Comparing with Eq.~(\ref{eq:photonsS}) we find
\beq
\label{eq:alphalambda}
\alpha_\Lambda = 2 {\cal{A}}_{\theta_s } = - \frac{1-|r|^2}{1+|r|^2},
\eeq
where we have noted the relation to the forward-backward asymmetry
${\cal{A}}_{\theta_s }$, (the analog of
${\cal{A}}_{\theta_S}$) for the angle $\theta_s$ between the
$\Lambda$ spin vector and
$\hat{p}_\Lambda $.  Our expression for $\alpha_\Lambda $ differs from
Refs.~\cite{mannelrecksiegel,huangyan98} by different
functions of baryon masses and from \cite{chuahehou} by an overall sign.
The variable $\alpha_\Lambda $ is determined by measuring the
angle $\theta_p $ in the $\Lambda$ rest frame between
the proton momentum vector from the
secondary decay $\Lambda \to p \pi^-$ and the direction parallel to
$\hat{p}_\Lambda$ or opposite to the $\Lambda_b$ momentum.
The distribution for this angle is proportional to
$(1 + \alpha_\Lambda \alpha \cos\theta_p )$, where $\alpha$ is the 
weak decay parameter for
$\Lambda \to p \pi^- $ which has been measured to high precision,
$\alpha = 0.642 \pm 0.013 $ \cite{pdg00}.
Thus, $\alpha_\Lambda $  can be related to the observable
forward-backward asymmetry in the angle $\theta_p$,
\beq
\label{eq:Athetap}
{\cal{A}}_{\theta_p} =
\frac{1}{2} \alpha_\Lambda \alpha = -\frac{\alpha}{2} \frac{1-|r|^2}{1+|r|^2} .
\eeq
It is apparent from Eqs.(\ref{eq:alambdab}) and (\ref{eq:Athetap}),
that both observables, ${\cal{A}}^\gamma$ and ${\cal{A}}_{\theta_p}$ can
only probe $|r|$. In Section \ref{sec:NLO} we show however, that at NLO in
$\alpha_s$ we are sensitive to direct CP violation in the decay
amplitudes and measurements of the CP averaged and flavor tagged
observables contain information beyond the magnitude of the
coupling ratio.

Although data indicate that the $b \to c$ vertex is
predominantly left-handed \cite{rizzoRHrevisited},
the possibility exists that New Physics could induce
tree level V+A currents.
In a SM based analysis of $b \to c \ell \nu_{\ell}$ mediated decays
a significant right-handed admixture would yield an effective
$\Lambda_b $ polarization that differs from its true value.
We have assumed here so far that this is not the case.
This hypothesis can itself be tested at a GigaZ facility
by comparing the value of $P_{\Lambda_b}$ extracted from
different measurements.
To be specific, {\it comparison of ${\cal{A}}^\gamma $ and
the $\Lambda$ polarization observable ${\cal{A}}_{\theta_p} $
provides an independent measurement of the $\Lambda_b$ polarization},
$P_{\Lambda_b } = \alpha {\cal{A}}^\gamma /{\cal{A}}_{\theta_p}$.
A discrepancy with the value of $P_{\Lambda_b} $ measured
in semileptonic $\Lambda_b \to \Lambda_c \ell \nu_{\ell} X$
decays would indicate the presence of non-standard
right-handed $b\to c$ currents.

Besides providing the above important consistency check we show in the next
section that
combining the measurements of ${\cal{A}}^\gamma $ and
${\cal{A}}_{\theta_p} $
has another advantage: a significant
increase in the statistical sensitivity to $|r|$.

\subsection{Sensitivity of the Observables ${\cal{A}}^\gamma $ and
${\cal{A}}_{\theta_p} $ to New Physics \label{subsec:sens}}

\begin{figure}[p]
\vskip -0.0truein
\hspace{4.2cm}(a) \hspace{7.8cm} (b) \hfill\mbox{ }
\centerline{\epsfysize=2.0in
{\epsffile{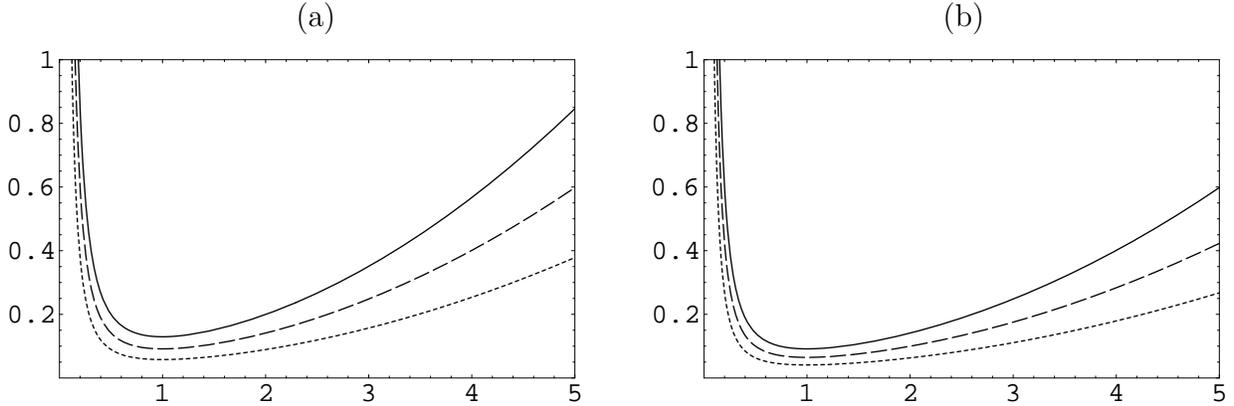}}}
\vskip -0.0truein
\caption[]{
(a) Relative statistical error in
$|r| = |C_{7}^\prime / C_{7} |$
as a function of $|r|$ obtained
from the angular asymmetry ${\cal{A}}^\gamma$ for
$2\cdot 10^9 $ (solid), $4 \cdot 10^{9}$ (long dashed), and $10^{10}$
(short-dashed) $Z$ bosons, corresponding to 760, 1520, and 3800 fully
reconstructed $\Lambda_b \to \Lambda \gamma$ decays, respectively,
given the efficiency estimates in the text and
${\cal{B}}(\Lambda_b \to \Lambda \gamma)= 7.5 \cdot 10^{-5}$.
(b)  Same as (a) but with twice the statistics, obtained by combining
${\cal{A}}^\gamma$ and
${\cal{A}}_{\theta_p}$.
At NLO, the figures give relative statistical errors in $|r^{\rm 
eff}|$ as a function of $|r^{\rm eff}|$.
Fig. (b) also gives the relative error in the ratio $r_{av}$ of CP 
even quantities, obtained from either 
${\cal{A}}^\gamma$ or ${\cal{A}}_{\theta_p}$.}
\label{fig:sensitivity}
\end{figure}

To illustrate the sensitivity of the angular asymmetry
${\cal{A}}^\gamma $ to the ratio $|r|$ we take
${\cal{B}} (\Lambda_b \to \Lambda \gamma) = 7.5 \cdot 10^{-5}$,
corresponding to approximately $2600$ $\Lambda_b \to \Lambda \gamma$
decays for $2 \cdot 10^9$ $Z$ bosons per year at a $Z$ factory.
We recall that the large theoretical uncertainty in the rate drops out
in ${\cal{A}}^\gamma $.
To estimate the number of fully reconstructed signal events
\footnote{We thank Su Dong for the reconstruction efficiency estimates.},
the total efficiency to reconstruct $\Lambda \to p \pi^-$ decays is
taken to be around 50\%, which includes acceptance losses,
tracking efficiency, and the
probability that the $\Lambda$ sometimes travels too far into the
central tracking system to leave much of a track when it decays.
In addition,
the efficiency for photon reconstruction is expected to be around
90\%.  Including the branching ratio of
${\cal{B}}(\Lambda \to p \pi^-)=0.639 \pm 0.005$ \cite{pdg00},
we obtain approximately $N = 760$ fully
reconstructed
signal events per year, ignoring cuts for background subtraction.
We further fix $P_{\Lambda_b}=-0.56$,
and do not take into account the experimental uncertainty from the boost.
The (absolute) statistical error in ${\cal{A}}^\gamma$ is
$\delta {\cal{A}}^\gamma = \sqrt{1- {{\cal{A}}^\gamma}^2}/\sqrt{N}$.
Our findings for the statistical sensitivity
are displayed in Fig.~\ref{fig:sensitivity}a for
1, 2 and 5 years of
running at design luminosity of $2 \cdot 10^{9}$ $Z$'s corresponding to $760$,
$1520$ and $3800$ fully reconstructed decays.

\begin{table}
          \begin{center}
          \begin{tabular}{|c|c|c|}
          \hline
          \multicolumn{1}{|c|}{ $ \# Z's$}   &
          \multicolumn{1}{|c|}{ ${\cal{A}}^\gamma$}      &
          \multicolumn{1}{|c|}{
          ${\cal{A}}^\gamma,{\cal{A}}_{\theta_p},CP$} \\
           \hline
$2 \cdot 10^{9} $ & $0.50 \leq |r| < 2.0$ \ ($0.34 \leq |r| < 2.9 $) &
$0.30 \leq |r| < 3.3$ \ ($0.23 \leq |r| < 4.4$)\\
$4 \cdot 10^{9} $ & $0.38 \leq |r| < 2.6$ \ ($0.28 \leq |r| < 3.6$) &
$0.25 \leq |r| < 4.0$ \ ($0.19 \leq |r| < 5.3$)  \\
$10 \cdot 10^{9}$ & $0.29 \leq |r| < 3.5$ \ ($0.21 \leq |r| < 4.7$) &
$0.19 \leq |r| < 5.2$ \ ($0.15 \leq |r| < 6.8$)\\
          \hline
         \end{tabular}
          \end{center}
\caption{ Ranges of the ratio $|r| = |C_{7}^\prime / C_{7} |$
that can be probed at $5 \sigma$ ($3 \sigma $ in parenthesis)
by measuring the angular asymmetry ${\cal{A}}^\gamma $ in
$\Lambda_b \to \Lambda \gamma$
decays (left column) for given number of $Z$'s.
In the right column, we combined
measurements from ${\cal{A}}^\gamma $ and
${\cal{A}}_{\theta_p} $ and  averaged over CP
conjugate decays. For details see Section \ref{subsec:sens}.
At NLO the left column corresponds to ranges of
$|r^{\rm{eff}}|$ obtained from ${\cal{A}}^\gamma $. In the
right column are shown the corresponding ranges for the ratio $r_{av}$
of CP even quantities, obtained from combined measurements of 
${\cal{A}}^\gamma $ and
${\cal{A}}_{\theta_p} $.
See Section \ref{sec:NLO} for details.}
\label{tab:ranges}
\end{table}

\begin{figure}[p]
\vskip -0.0truein
\centerline{\epsfysize=2.0in
{\epsffile{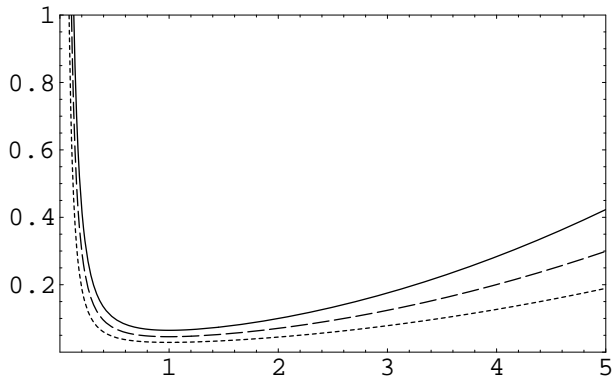}}}
\vskip -0.0truein
\caption[]{
Relative statistical error in
$|r| = |C_{7}^\prime / C_{7} |$ as a function of $|r|$
extracted from the angular asymmetry ${\cal{A}}^\gamma$ for
$2\cdot 10^9 $ (solid), $4 \cdot 10^{9}$ (long dashed), and $10^{10}$
(short-dashed) $Z$ bosons, corresponding to 760, 1520, and 3800 fully
reconstructed $\Lambda_b \to \Lambda \gamma$ decays, respectively,
obtained by combining the values of $|r|$ extracted from ${\cal{A}}^\gamma$ and
${\cal{A}}_{\theta_p}$ and
averaging over
CP conjugate decays, in the limit of no CP violation.
At NLO, the figure gives the relative statistical error in the ratio $r_{av}$
of CP even quantities, obtained by combining ${\cal{A}}^\gamma$ and
${\cal{A}}_{\theta_p}$.
See Section \ref{sec:NLO} for details.}
\label{fig:sensitivitycombined}
\end{figure}

Comparing the expressions for ${\cal{A}}^\gamma $ and
${\cal{A}}_{\theta_p} $
in Eqs.~(\ref{eq:alambdab}) and (\ref{eq:Athetap}),
it is clear that for comparable
magnitudes of $\alpha$ and $P_{\Lambda_b}$ as indicated by
HQET and LEP measurements, the
statistical sensitivities of the two observables to $|r|$ are similar.
Furthermore, there should not be a significant additional uncertainty
in ${\cal{A}}_{\theta_p} $
due to the extra boost from the $\Lambda_b $ to $\Lambda$ rest frames,
since these decays are fully reconstructed.
In Fig.~\ref{fig:sensitivity}b we show the sensitivity obtained from combined
measurements of ${\cal{A}}^\gamma $ and
${\cal{A}}_{\theta_p} $.
Finally, by the time that the GigaZ will be in operation we will
already know from the $B$ factories whether or not
there is significant direct CP violation in $b \to s \gamma$ mediated
decays.  In the limit of none or very little CP violation like in the SM,
${\cal{A}}^\gamma$ and ${\cal{A}}_{\theta_p}$ are CP-even or close to it.
In this case, we can roughly quadruple the statistical power by
combining the measurements of
$|r|$ extracted from  ${\cal{A}}^\gamma $ and ${\cal{A}}_{\theta_p} $
and averaging over the CP conjugate decays.
This possibility is illustrated in Fig.~\ref{fig:sensitivitycombined}.
The ranges for $|r|$ that can be probed would be substantially increased
as demonstrated in Table \ref{tab:ranges}.
Here we show, for comparison,
the $5\sigma$ ranges ($3 \sigma $ in parenthesis) obtained from
analyzing ${\cal{A}}^\gamma$
alone, and those obtained by combining with ${\cal{A}}_{\theta_p}$
and including
both the $\Lambda_b$ and CP conjugate decays.

We return to the issue of CP violation below,
and show that even with sizeable CP violation
the CP averaged observables are useful, yielding information
on the CP even part of an effective coupling ratio rather than on $|r|$.

\section{NLO Considerations and CP Violation \label{sec:NLO}}

In this section we estimate next-to-leading order (NLO) effects in
$\Lambda_b \to \Lambda \gamma$ decays, where use
is made of the corresponding results
for inclusive $B \to X_s \gamma$ decays \cite{misiak,GHW}.
An important addition to the LO analysis is the sensitivity to
CP violation at ${\cal O} (\alpha_s )$.
In the following sections we give the matrix element for
$\Lambda_b \to \Lambda \gamma$ decays at NLO,
discuss CP violating effects, and
work out the relations between the coefficients appearing in the
modified matrix element and
the observables defined in Section \ref{sec:asymmetry}.

\subsection{The $\Lambda_b \to \Lambda \gamma$ Matrix Element
at ${\cal O} (\alpha_s )$ \label{subsec:matrix}}

As already mentioned in Section \ref{sec:asymmetry}
helicity changing long distance effects
are expected to alter the $\Lambda_b \to \Lambda \gamma$
amplitude \cite{mannelrecksiegel,LDkstar,pirjol} and therefore the
angular asymmetry observables ${\cal{A}}^\gamma $ and ${\cal{A}}_{\theta_p} $
by at most a few percent.
In the following we ignore these effects, but will allow
for contributions from hard gluon exchanges beyond leading order.
The $\Lambda_b \to \Lambda \gamma$ amplitude
can be parametrized in terms of effective coefficients $D,D^\prime$ as
\beq
\label{eq:matrix}
A (\Lambda_b \to \Lambda \gamma )= -\frac{G_F }{\sqrt{2}} V_{t s}^\ast  V_{tb}
(D  \langle \Lambda \gamma | Q_7 | \Lambda_b \rangle  + D^\prime
\langle \Lambda \gamma | Q^\prime_7 | \Lambda_b \rangle ), \eeq
where $\langle Q_7 \rangle $, $\langle Q_7^\prime \rangle $ are the
leading-order matrix elements following from Eq.~(\ref{eq:sigma}).
To ${\cal{O}}(\alpha_s)$
\beqa
\label{eq:Ceff}
D & = & C_{7}^{(0)} + \frac{\alpha_s}{4 \pi}
(C_{7}^{(1)} + C_2^{(0)} k_2 +
C_{8}^{(0)} k_8) \nonumber \\
D^\prime & = &  C_{7}^{\prime (0)} + \frac{\alpha_s}{4 \pi}
(C_{7}^{\prime (1)} +C_{8}^{\prime (0)} k_8) .
\eeqa
Here, the coefficients $k_i $ account for the ${\cal O} (\alpha_s )$ matrix
elements of the operators $Q_i^{(\prime)}$ and include
CP conserving strong phases.
As usual, $Q_2 = (\bar{c} \gamma_\mu L b) ( \bar{s} \gamma^\mu L c)$ is the
current-current operator and
$Q^{(\prime)}_8$ is the chromomagnetic dipole operator analog of
$Q^{(\prime)}_7$, see e.g.~\cite{heffbig}.
We have further assumed that the flipped current-current operator
${\cal{O}}_2^{\prime}=(\bar{c} \gamma_\mu R b)
(\bar{s} \gamma^\mu R c)$ is of negligible strength and does not
contribute to the $\alpha_s$-corrected
matrix element.
The superscripts $(0)$ and $(1)$ denote
LO and NLO  contributions to the Wilson coefficients, respectively.

The coefficients $k_i $ receive contributions from gluonic loops in the
$b \to s \gamma$ transition  \cite{misiak,GHW} as wells as
from hard interactions with the spectator quarks.
Studies for exclusive $B \to K^* \gamma$ decays have shown that
$\alpha_s$-corrections from diagrams involving spectator quarks are
smaller than those without spectator interactions
\cite{Bosch:2001gv,Beneke:2001gv}.
Thus, while an explicit NLO calculation for
$\Lambda_b \to \Lambda \gamma$ decays would be desirable,
existing calculations of the $k_i$ for inclusive $b \to s \gamma$ decays
should provide an estimate of the dominant NLO effects to
the exclusive decay.
Note that in Eq.~(\ref{eq:Ceff}) we have absorbed the
'factorizable' vertex correction of the operators $ Q_7, Q_7^\prime$
into the form factor \cite{Bosch:2001gv,Beneke:2001gv,benekefeldmann00}.

We will allow for weak CP violating phases in the
Wilson coefficients of the operators
$Q_7^{(\prime)}$ and $Q_8^{(\prime)}$.
The effective coefficients $\bar{D} $ and $\bar{D}^{\prime}$
for the CP conjugate decay $\bar{\Lambda}_b \to \bar{\Lambda} \gamma $
are then obtained by replacing
$C_{7}^{(0),(1)}$, $C_{8}^{(0)}$ and their primed counterparts
in (\ref{eq:Ceff}) with their complex conjugates.

\subsection{Direct CP Violation \label{sec:CP}}

Direct CP violating effects can arise at
${\cal{O}}(\alpha_s)$ from interference
between the weak and strong phases in the decay amplitudes,
for example inducing a non-zero asymmetry
in the decay rates:
\begin{equation}
a_{CP}^{\Lambda_b}\equiv
\frac{\Gamma-\bar{\Gamma}}{\Gamma+\bar{\Gamma}},
\label{eq:acp}
\end{equation}
where $\Gamma,~\bar{\Gamma}$ denote the
total decay rates for $\Lambda_b \to \Lambda \gamma$ and the CP conjugate
mode, respectively.
In general, the CP asymmetry in $b\to s$ transitions is CKM suppressed in
scenarios which only contain the weak CKM phase of the SM:
$a_{CP} \sim  \alpha_s (m_b) {\rm{Im}}[V_{us}^* V_{ub}/V_{ts}^* V_{tb}] =
\alpha_s (m_b) \lambda^2 \eta $, where
$\lambda$ and $\eta$ are Wolfenstein parameters, and
$\lambda^2 \eta \sim 0.02$.
We estimate $a_{CP}^{\Lambda_b} \leq {\cal{O}}(1)\% $ in the SM
from calculations of
the inclusive $B \to X_s \gamma $ \cite{kaganneubertCP}
or exclusive $B \to K^* \gamma $ \cite{Bosch:2001gv,GSW} rate asymmetries
and neglect such small CKM induced effects below.
However, new CP violating contributions to $Q_7^{(\prime)} $ or
$Q_8^{(\prime)}$ can give rise to sizable effects.
In particular, it has been shown that
CP violating rate asymmetries of order 10\% or larger are possible
for inclusive $B \to X_s \gamma$ decays in a variety of New Physics models
\cite{kaganneubertCP}, so that similarly large values for
exclusive asymmetries can be expected.
Experimentally, the current best bound is given as
$a_{CP}(B \to K^* \gamma)=-0.035 \pm 0.076 \pm 0.012$
\cite{babarLP01}, whereas
inclusive CP asymmetries are not very constrained yet
$-0.27 < a_{CP}(B \to X_s \gamma) < +0.10$ at $90 \%$ C.L.~\cite{cleoAcp}.

Below we will discuss the angular asymmetries beyond leading
order, allowing in general for CP violating
effects.  We will see that by combining measurements of these
observables with branching ratio measurements, for the CP
conjugate decay modes, it will be possible
to determine the CP odd and CP even components of both the SM and
opposite chirality contributions
to the $\Lambda_b \to \Lambda \gamma$ decay rate.

\subsection{The Observables at NLO}

In the CP conserving limit the angular asymmetry observables
${\cal{A}}^\gamma$ and ${\cal{A}}_{\theta_p}$ are CP-even, i.e.,
${\cal{A}}^\gamma  = \bar{{\cal{A}}}^\gamma $,
and ${\cal{A}}_{\theta_p}  = \bar{{\cal{A}}}_{\theta_p} $,
where
${\cal{A}}$ and $\bar{{\cal{A}}}$ are the observables for
the $\Lambda_b$ and CP conjugate $\bar{\Lambda}_b$ decays, respectively.
However,
the angular asymmetries of the CP conjugate modes
will in general differ at next-to-leading order and higher
if there are new CP violating contributions to $Q_7^{(\prime)}$ or
$Q_8^{(\prime)}$.
We parametrize the angular asymmetry observables as
\beqa
\label{eq:CPconjugateobs}
{\cal{A}}^\gamma & = & - {P_{\Lambda_b} \over 2}
\frac{1-|r^{\rm{eff}}|^2}{1+|r^{\rm{eff}}|^2},~~~~~{\cal{A}}_{\theta_p}
=  - {\alpha \over 2}
\frac{1-|r^{\rm{eff}}|^2}{1+|r^{\rm{eff}}|^2} , \nonumber \\
{\bar{\cal{A}}}^\gamma & = & - {P_{\Lambda_b} \over 2}
\frac{1-|\bar{r}^{\rm{eff}}|^2}{1+|\bar{r}^{\rm{eff}}|^2},
~~~~~{\bar{\cal{A}}}_{\theta_p}  =  - {\alpha \over 2}
\frac{1-|\bar{r}^{\rm{eff}}|^2}{1+|\bar{r}^{\rm{eff}}|^2},
\eeqa
where the effective ratios are defined as
\beq
r^{\rm{eff}} \equiv D^{\prime } /D,~~~~
\bar{r}^{\rm{eff}} \equiv \bar{D}^{\prime} /\bar{D} .
\label{eq:reff}
\eeq
Thus the flavor specific angular asymmetry observables
actually probe the effective ratios $|r^{\rm{eff}} |$
and $|\bar{r}^{\rm{eff}} |$, rather than
the ratio of short distance Wilson coefficients $|C_{7}^\prime / C_{7} |$.
It is straightforward to carry over the results obtained
in Section \ref{sec:asymmetry} for the
experimental sensitivity to $|r|$:  The ranges in
$|r^{\rm{eff}} |$ ($|\bar{r}^{\rm{eff}} |$)
that can be probed by measuring ${\cal{A}}^\gamma$
($\bar{{\cal{A}}}^\gamma$) can be read off from
Fig.~\ref{fig:sensitivity}a.
Measurements of ${\cal{A}}_{\theta_p}$ will give a similar reach
since we assumed
that $P_{\Lambda_b}$ is of similar magnitude to the
$\Lambda$ decay parameter $\alpha$. The sensitivity
that can be expected by combining measurements of the two observables is given
approximately in Fig.~\ref{fig:sensitivity}b.

In the following it is  convenient to separate
$|D^{(\prime) }|^2 $ into CP even and CP odd
components, denoted by $|D^{(\prime) }|^{2  +}$ and $|D^{(\prime)
}|^{2  -}$, respectively, such that
\beq
|D^{(\prime) }|^{2 } = |D^{(\prime) }|^{2  +} + |D^{(\prime) }|^{2 -},~~~~~
|\bar{D}^{(\prime) }|^{2 } = |D^{(\prime) }|^{2  +} -
|D^{(\prime)}|^{2 -}.\nonumber
\eeq
At next-to-leading order we obtain
\beqa
|D|^{2  +} & = & |C_{7}^{(0)} |^2 + \frac{\alpha_s }{2 \pi}
( {\rm{Re}}[C_{7}^{(0)} C_{7}^{(1) *} ]
+ {\rm{Re}}[C_{7}^{(0)} C_{2}^{(0) *} ] {\rm{Re}}  k_2
+ {\rm{Re}}[C_{7}^{(0)} C_{8}^{(0) *} ] {\rm{Re}}  k_8) , \nonumber \\
|D|^{2  -}  & = & \frac{\alpha_s }{2 \pi}
( {\rm{Im}}[C_{7}^{(0)} C_{2}^{(0) *} ] {\rm{Im}} k_2
+ {\rm{Im}}[C_{7}^{(0)} C_{8}^{(0) *} ] {\rm{Im}}  k_8), \nonumber \\
|D^{\prime }|^{2  +} & = & |C_{7}^{\prime (0)} |^2
+ \frac{\alpha_s}{2 \pi}
({\rm{Re}}[C_{7}^{\prime (0)} C_{7}^{ \prime (1) *} ] +
{\rm{Re}}[C_{7}^{\prime (0)} C_{8}^{\prime (0) *} ] {\rm{Re}}  k_8),
\nonumber \\
|D^{\prime }|^{2  -}  & = & \frac{\alpha_s }{2 \pi}
{\rm{Im}}[C_{7}^{\prime (0)} C_{8}^{\prime (0) *} ] {\rm{Im}}  k_8.
\label{eq:Xnlo}
\eeqa
Note that the CP odd components $|D|^{2  -} $ and
$|D^{\prime }|^{2-} $ arise only at ${\cal{O}}(\alpha_s )$.
There are three CP even observables: the averages
over the CP conjugate modes of the branching ratio and of the angular
asymmetry observables, denoted
${\cal{B}}_{av} $, ${\cal{A}}^\gamma_{av} $ and
${\cal{A}}_{\theta_p}^{av} $, respectively.
The three corresponding CP odd observables are the rate asymmetry
$a_{CP}^{\Lambda_b}$
and the angular asymmetry differences
${\cal{A}}^\gamma - \bar{\cal{A}}^\gamma $ and
${\cal{A}}_{\theta_p} - \bar{\cal{A}}_{\theta_p} $.

All four components $|D^{(\prime) }|^{2  +} $ and
$|D^{(\prime) }|^{2 -}$ can in principle be uniquely determined from experiment
via the relations
\beqa
{\cal{B}}_{av} & = & \tau(\Lambda_b)
\Gamma_0
(|D|^{2  +} + |D^{\prime }|^{2  +}) ,  \\
a_{CP}^{\Lambda_b} & = &
\frac{|D|^{2  -} + |D^{\prime }|^{2  -}}{|D|^{2  +} + |D^{\prime }|^{2
+}} , \label{eq:Acp} \\
{\cal{A}}^\gamma_{av} +  a_{CP}^{\Lambda_b} \frac{{\cal{A}}^\gamma -
\bar{\cal{A}}^\gamma }{2} & = & - \frac{P_{\Lambda_b}}{2}
\frac{|D|^{2  +} - |D^{\prime }|^{2  +} }
{|D|^{2  +} + |D^{\prime }|^{2  +}}  , \label{eq:Aav} \\
\frac{{\cal{A}}^\gamma - \bar{\cal{A} }^\gamma}{2} +
a_{CP}^{\Lambda_b} {\cal{A}}^\gamma_{av}  & = &
- \frac{P_{\Lambda_b}}{2}
\frac{|D|^{2  -} - |D^{\prime }|^{2  -} }
{|D|^{2  +} + |D^{\prime }|^{2  +}}  , \label{eq:Xrelns}
\eeqa
plus two equations involving
the $\Lambda $ polarization observables ${\cal{A}}_{\theta_p}$
and $\alpha$, obtained
by substituting for ${\cal{A}}^\gamma$ and $P_{\Lambda_b}$,
respectively, in the last two equations above.  Note that the second term
on the l.h.s. of Eq. (\ref{eq:Aav}) first enters at order
$\alpha_s^2 $ and should be neglected in a NLO analysis.

An important result following immediately from Eq.~(\ref{eq:Aav}) is
that the CP
averaged angular observables
${\cal{A}}^\gamma_{av}$ and ${\cal{A}}_{\theta_p}^{av}$ in general
determine the
ratio of CP even quantities
\beq
r_{av} \equiv \sqrt{\frac{|D^{\prime }|^{2  +}}{|D|^{2  +}} },
\nonumber
\eeq
at NLO via equations analogous to Eqs.~(\ref{eq:alambdab}) and
(\ref{eq:Athetap}), respectively.
Furthermore, the full statistical reach of a GigaZ facility, as discussed in
Section \ref{subsec:sens}, is available
since both the $\Lambda_b $ and $\bar{\Lambda}_b$ decays are included in
CP averaged
quantities:
The sensitivity to $r_{av}$
that could be obtained from measurements of either ${\cal{A}}^\gamma_{av}$ or
${\cal{A}}_{\theta_p}^{av}$ can be read off from
Fig.~\ref{fig:sensitivity}b,
whereas the sensitivity for a combined analysis
is given in Fig.~\ref{fig:sensitivitycombined},
also see Table \ref{tab:ranges}.
A non-zero measurement of $r_{av}$
would be a clean signal for
New Physics with non-standard chirality structure,
given that in the SM $r_{av} \sim m_s /m_b $.

\subsection{Estimates of NLO Effects}

Small measured values for the CP violating rate asymmetry, $a_{CP}^{\Lambda_b}$
would generally imply that $|D|^{2 -} $ and  $|D^\prime|^{2 -} $, and therefore
${\cal{A}}^\gamma - \bar{{\cal{A}}}^\gamma$ and
${\cal{A}}_{\theta_p} - \bar{\cal{A}}_{\theta_p} $ are small.
This can be seen explicitly
from Eqs.~(\ref{eq:Acp}) and (\ref{eq:Xrelns}).
Furthermore, if $|D^\prime|^{2 -} = 0$, i.e., if the New Physics
contributions to $C_{7}^\prime $
and $C_{8}^\prime $ have a common weak phase, then
\beqa
\label{eq:AgammaCP}
{\cal{A}}^\gamma - \bar{{\cal{A}}}^\gamma
= - 2 a_{CP}^{\Lambda_b} P_{\Lambda_b}
\frac{|D^\prime |^{2 +}}{|D |^{2 +} + |D^\prime |^{2 +}},~~~~
{\cal{A}}_{\theta_p} - \bar{\cal{A}}_{\theta_p}
= - 2 a_{CP}^{\Lambda_b} \alpha
\frac{|D^\prime |^{2 +}}{|D |^{2 +} + |D^\prime |^{2 +}},
\eeqa
where the equalities hold up to and including terms of ${\cal{O}}(\alpha_s^2)$.
Setting $|D^\prime|^{2 -} = 0$ would be a good approximation if there
were a single dominant New
Physics source, such as the virtual exchange of a new heavy
particle, contributing to both the
magnetic and chromomagnetic dipole operators.
In such models an upper bound is obtained on the angular CP asymmetries,
$|{\cal{A}}^\gamma - \bar{{\cal{A}}}^\gamma |
< 2 |a_{CP}^{\Lambda_b} P_{\Lambda_b}| $, and
$|{\cal{A}}_{\theta_p} - \bar{{\cal{A}}}_{\theta_p}|
< 2 |a_{CP}^{\Lambda_b}| \alpha $.
Barring large accidental cancellations, data on
$a_{CP} (B \to X_s \gamma)$ or
$a_{CP} (B \to K^* \gamma)$ may serve here as a first estimate,
so roughly $|{\cal{A}}^\gamma - \bar{{\cal{A}}}^\gamma |,
|{\cal{A}}_{\theta_p} - \bar{{\cal{A}}}_{\theta_p}| \lsim
{\cal{O}}(10\%) $, using the experimental information given in Section
\ref{sec:CP}

Finally, we ask by how much
$r_{av}$ and $|r^{\rm {
eff}} |$ could
differ from the
leading order ratio $|C_{7}^{\prime (0)} / C_{7}^{(0)} |$, which was
the focus of the previous sections.
At NLO order we have
\beqa
\label{eq:Xratio}
r_{av} & = &
\frac{|C_{7}^{\prime (0)}|}{|C_{7}^{(0)} |}
\left( 1 + \frac{\alpha_s}{4 \pi }
( {\rm{Re}} [ \frac{C_{7}^{\prime (1)} }{C_{7}^{\prime (0)} }
- \frac{C_{7}^{(1)} }{C_{7}^{(0)} } ] +
{\rm{Re}}  k_8 {\rm{Re}}  [ \frac{C_{8}^{\prime (0)} }{C_{7}^{\prime (0)} }
- \frac{C_{8}^{(0)} }{C_{7}^{ (0)} } ] - {\rm{Re}}  k_2 {\rm{Re}} [
   \frac{C_{2}^{ (0)} }{C_{7}^{ (0)} }] ) \right), \nonumber \\
|r^{\rm { eff}} | & =& r_{av}
+\frac{|C_{7}^{\prime (0)}|}{|C_{7}^{(0)} |} \frac{\alpha_s}{4 \pi }
(
{\rm{Im}}  k_8 {\rm{Im}}  [ \frac{C_{8}^{ (0)} }{C_{7}^{ (0)} }
- \frac{C_{8}^{ \prime (0)} }{C_{7}^{\prime (0)} } ] +
{\rm{Im}}  k_2 {\rm{Im}} [
   \frac{C_{2}^{ (0)} }{C_{7}^{ (0)} }] ).
\eeqa
As discussed at the beginning of Section \ref{subsec:matrix},
an estimate of the ${\cal O}(\alpha_s ) $ matrix element
can be obtained from inclusive $b \to s \gamma $ decays
keeping only the finite virtual corrections.
The exclusive coefficients $k_i$ can be
roughly approximated by the corresponding inclusive ones
\cite{misiak,GHW,kaganneubert}, yielding
\beqa
\label{eq:kiestimates}
{\rm{Re}}  k_2 & \approx &  -4.09 + 12.78 (\frac{m_c}{m_b} -0.29) +
\frac{416}{81} {\rm{ln}}  \frac{m_b}{\mu}, \nonumber \\
{\rm{Im}}  k_2 & \approx &  -0.45+5.18(\frac{m_c}{m_b} -0.29) , \nonumber \\
{\rm{Re}}   k_8 & \approx & \frac{44}{9} -\frac{8 \pi^2 }{27}
-\frac{32}{9} {\rm{ln}}  \frac{m_b}{\mu} ,
   ~~~~~~~~{\rm{Im}}  k_8 \approx \frac{8 \pi}{9} , \nonumber
\eeqa
where $\mu $ is the renormalization scale.
Taking $C_7^{(0)} $ in Eq.~(\ref{eq:Xratio}) to be approximately equal to
the SM value,
and allowing $\mu$ to vary between $m_b /2 $ and $m_b$,
we find that the ${\cal O}(\alpha_s )$ corrections to
$r_{av}$ or $|r^{\rm{eff}} |$
induced by the
matrix element of $Q_2 $ are of order $5\% -20\%$.
Shifts due to the matrix elements of
$Q_8 $, $Q_8^\prime$ would be of order 1\% if
$C_8 \sim C_7 $ and $C_7^\prime \sim C_8^\prime $, as in the SM.
However, in models with enhanced chromomagnetic dipole
operators the correction could again be
of order 10\%.
Therefore, although measurements of the observables associated with
$\Lambda_b \to \Lambda \gamma$
could give unambiguous evidence for New Physics with non-SM chirality,
it will be difficult to obtain precision constraints
on the underlying short-distance contributions to the dipole operators
in the absence of a first-principles calculation
of the  coefficients $k_i$ in exclusive $\Lambda_b \to \Lambda \gamma$
decays.

\section{Conclusions and Outlook \label{sec:conclusions}}

We have studied the radiative decay
$\Lambda_b \to \Lambda \gamma$  as a probe of New Physics.
A novel observable was proposed
which makes use of the polarization of $\Lambda_b$ baryons produced at the $Z$:
the angular asymmetry of the photon momentum with respect to the
$\Lambda_b$ boost axis.
We have also considered the angular asymmetry associated with the secondary
decays $\Lambda \to p \pi^- $.
The two observables are sensitive to the ratio
$C_{7}^\prime /C_{7}$ of opposite chirality to Standard Model chirality
$b \to s \gamma $ Wilson coefficients.
In the Standard Model this ratio is only a few percent but can be
sizeable in many of its extensions e.g.~the MSSM beyond minimal flavor
violation.
Statistical sensitivities
to this ratio were worked out, including reconstruction efficiency estimates.
Our findings are compiled in Table \ref{tab:ranges} and in three
figures for the case of a proposed GigaZ facility \cite{tesla,usaLC}
with $\approx 2 \times 10^{9}$ $Z$'s per year.
Wide ranges of $C_{7}^\prime /C_{7}$ are
accessible to experimental study of angular asymmetries in
$\Lambda_b \to \Lambda \gamma$ decays, allowing a clear separation from
the SM prediction.

In addition to the search for non-standard chiralities, one can
probe for non standard CP phases in $\Lambda_b \to \Lambda \gamma$ decays,
if a flavor tagged analysis of angular asymmetries and branching ratios
is performed.
In general, at NLO and allowing for direct CP violation, four
independent contributions enter the $\Lambda_b \to \Lambda \gamma$
and CP conjugate decay widths: CP even and CP odd, each with SM and
opposite chiralities.
All four can, in principle, be determined from such an analysis.
An important result is that the CP
averaged angular observables, which have the greatest statistical
reach, determine the
relative strengths of the CP even contributions with opposite and 
Standard Model
chiralities, generalizing the
leading order dependence on $C_{7}^\prime /C_{7}$.  A non-zero
measurement of this ratio
would provide a clean signal for New Physics.

Parts of the analysis presented here, namely
measurements of rates and
the $\Lambda$ decay polarization observable, including studies of CP
violation do not require
polarized $\Lambda_b$'s and can be carried out
at hadron colliders like the Tevatron and the LHC.
It might also be worthwhile to explore the possibility of heavy
baryon production with sufficient polarization in a hadronic environment
e.g.~with polarized beams.

To estimate the total $\Lambda_b \to \Lambda \gamma$ rate,
we derived form factor relations
for heavy-to light baryons decays in the large energy limit.
This allows us to relate the form factors to existing estimates
derived using non
perturbative methods and data.
We emphasize that the relations we have worked out are
useful for many other heavy-to-light decays at large recoil.
In particular,
we have shown that the zero of the dilepton forward-backward asymmetry
in $\Lambda_b \to \Lambda \ell^+ \ell^-$ decays
is independent of form factors to lowest order in the large energy expansion.
The form factor relations are also necessary for predicting
the proton angular asymmetry in polarized $\Lambda_b \to p \ell
\nu_\ell $ decays,
which provides an important test of the $V-A$ structure of the
$b \to u$ charged current at a GigaZ facility.

We stress the importance of a precise measurement of
the $\Lambda_b$ polarization from semileptonic
$\Lambda_b \to \Lambda_c \ell \bar{\nu}_\ell X$ decays at the GigaZ; a
significant improvement on the LEP measurements will be required.
Comparison with the polarization extracted from the angular
asymmetries in $\Lambda_b \to \Lambda \gamma$ provides a
consistency check of the $V-A$ structure of the $b \to c$ charged current.
The latter should also be testable via angular asymmetries
in exclusive $\Lambda_b \to (\Lambda_c \to \Lambda \pi, \Sigma \pi)
\ell \bar{\nu}_\ell $
decays.

It is promising to extend the study presented here to the
semileptonic decays
$\Lambda_b \rightarrow \Lambda \ell^+ \ell^-$
and  $\Lambda_b \rightarrow \Lambda \nu \bar{\nu}$,
with Standard Model branching ratios in the interesting
range of $10^{-5}$ to $10^{-6}$.
In a companion paper
\cite{lambdaphi} we discuss rare hadronic
two-body decays, focusing on the decay
$\Lambda_b \rightarrow \Lambda \phi $, which is estimated to have
a Standard Model branching ratio of a
few times $10^{-5}$. This decay offers a unique
sensitivity
to the chirality structure of four-quark `penguin{'} (see e.g.~\cite{heffbig})
operators.  The decays $\Lambda_b \to \Lambda \pi, \Lambda \rho$ are also
interesting since they violate isospin, thus providing a probe of the
electroweak
penguin operators.
Finally, certain
hadronic 2-body decays can explore the origin and limitations of the
factorization hypothesis \cite{diehlhiller}.
All of this should be part of a
rich and unique $b$-physics program at a future high luminosity $Z$
factory.

\noindent
{\bf Acknowledgements}
We would like to thank Gerhard Buchalla,
Stan Brodsky, Markus Diehl, Su Dong, Michael Peskin and Tom Rizzo
for useful and stimulating discussions and communications.  G.H. 
would like to thank
the CERN and DESY theory groups, where part of this work was carried 
out, for their hospitality,
and A.K. would like to thank the SLAC and DESY theory groups for 
their hospitality.

\end{document}